\documentclass[aps,twocolumn,preprintnumbers,amsmath,amssymb,superscriptaddress,floatfix,nofootinbib]{revtex4}

\usepackage{lipsum}
\usepackage{graphicx}
\usepackage{epsfig}
\usepackage{epstopdf}
\usepackage{hyperref}
\usepackage{amsmath}
\usepackage{amsfonts}
\usepackage{amssymb}

\begin{document}

\title{Revising the $f_1(1420)$ resonance}

\author{V.~R.~Debastiani}
\email{vinicius.rodrigues@ific.uv.es}
\affiliation{Departamento de
F\'{\i}sica Te\'orica and IFIC, Centro Mixto Universidad de
Valencia-CSIC Institutos de Investigaci\'on de Paterna, Aptdo.
22085, 46071 Valencia, Spain}

\author{F.~Aceti}
\email{aceti@ific.uv.es}
\affiliation{Departamento de
F\'{\i}sica Te\'orica and IFIC, Centro Mixto Universidad de
Valencia-CSIC Institutos de Investigaci\'on de Paterna, Aptdo.
22085, 46071 Valencia, Spain}

\author{Wei-Hong~Liang}
\email{liangwh@gxnu.edu.cn}
\affiliation{Departamento de
F\'{\i}sica Te\'orica and IFIC, Centro Mixto Universidad de
Valencia-CSIC Institutos de Investigaci\'on de Paterna, Aptdo.
22085, 46071 Valencia, Spain}
\affiliation{Department of Physics, Guangxi Normal University,
Guilin 541004, China}

\author{E.~Oset}
\email{oset@ific.uv.es}
\affiliation{Departamento de
F\'{\i}sica Te\'orica and IFIC, Centro Mixto Universidad de
Valencia-CSIC Institutos de Investigaci\'on de Paterna, Aptdo.
22085, 46071 Valencia, Spain}

\date{\today}

\begin{abstract}
We have studied the production and decay of the $f_1(1285)$ into $\pi a_0(980)$ and $K^* \bar K$ as a function of the mass of the resonance and find a shoulder around 1400 MeV, tied to a triangle singularity, for the $\pi a_0(980)$ mode, and a peak around 1420 MeV with about 60 MeV width for the $K^* \bar K$ mode. Both these features agree with the experimental information on which the $f_1(1420)$ resonance is based. In addition, we find that if the $f_1(1420)$ is a genuine resonance, coupling mostly to $K^* \bar K$ as seen experimentally, one finds unavoidably about a 20\% fraction for $\pi a_0(980)$ decay of this resonance, in drastic contradiction with all experiments. Altogether, we conclude that the $f_1(1420)$ is not a genuine resonance, but the manifestation of the $\pi a_0(980)$ and $K^* \bar K$ decay modes of the $f_1(1285)$ at higher energies than the nominal one.
\end{abstract}

\maketitle

\section{Introduction}

The $f_1(1420)$ resonance is catalogued in the Particle data book \cite{PDG} as an $I^G(J^{PC}) = 0^+(1^{++})$ state and has been observed in over 20 experiments. Its mass is $M = 1426.4 \pm 0.9$ MeV and its width $\Gamma = 54.9 \pm 2.6$ MeV. Its dominant decay mode is $K \bar K^*$. In Ref. \cite{Armstrong:1984rn} 100\% of the width is associated to the $K \bar K^* + \mathrm{c.c.}$ mode. In Ref. \cite{Armstrong:1992hm} other modes are also searched for, with negative results, concluding again that the $K \bar K^* + \mathrm{c.c.}$ channel exhausts the decay width. In Ref. \cite{Achard:2007hm} the authors also conclude that the $f_1(1420)$ decays into $K \bar K^* + \mathrm{c.c.}$ 100\%. In Ref. \cite{Barberis:1998by} the decay mode $\pi a_0(980)$ is reported with $\Gamma(\pi a_0(980))/\Gamma(K \bar K^* + \mathrm{c.c.}) = 0.04\pm0.01\pm0.01$.
In this latter paper a clean peak is seen for the $f_1(1285)$ in the $\pi a_0(980)$ mode, followed by a broader structure around 1400 MeV with much smaller strength, that is tentatively associated to the $f_1(1420)$, with the comment ``The shoulder at 1.4 GeV can be interpreted as an $a_0(980) \pi$ decay mode of the $f_1(1420)$'' with no devoted work to support this assertion. A discussion on mesons in the 1400 MeV region can be seen in the PDG review \cite{PDG:Amsler}.

In the present work we shall provide a different explanation of the experimental findings, showing that the $K^* \bar K$ peak associated to the $f_1(1420)$ is the manifestation of the $K \bar K^* + \mathrm{c.c.}$ decay mode of the $f_1(1285)$. On the other hand, the broad peak for $\pi a_0(980)$ decay in the region of 1400 MeV will be explained as a consequence of a triangle singularity, due to $f_1(1285) \to K^* \bar K ,~ K^* \to \pi K,~ K \bar K \to a_0(980)$.
The $K^* \bar K$ decay mode of the $f_1(1285)$ appears as ``not seen'' in the PDG \cite{PDG}. Indeed the $f_1(1285)$ is 100 MeV below the $K^* \bar K$ threshold. However, the $K \bar K \pi$ mode is reported with a branching fraction of 9\%. These features found an adequate answer in several works \cite{Aceti:2015zva,Aceti:2015pma}, where the $f_1(1285)$ was considered as a dynamically generated resonance. This state, together with all the low-lying axial vector resonances, were obtained in Refs. \cite{Lutz:2003fm,Roca:2005nm} as dynamically generated states from the interaction of pseudoscalar mesons with vector mesons, using a coupled channels unitary scheme with chiral dynamics for the meson interaction \cite{Birse:1996hd}. In the particular case of the $f_1(1285)$, the $K \bar K^* + \mathrm{c.c.}$ is the single channel in the coupled channel approach \cite{Roca:2005nm}.
The work in Refs. \cite{Lutz:2003fm,Roca:2005nm}, using the lowest order chiral Lagrangian, has been extended in Ref. \cite{Zhou:2014ila} including higher order terms, but in the case of the $f_1(1285)$ the higher order terms were found essentially negligible.

In this picture for the $f_1(1285)$, a good description of $\pi a_0(980)$ and the isospin forbidden $\pi f_0(980)$ decay modes were well reproduced \cite{Aceti:2015zva}.
Actually the $\pi f_0(980)$ decay mode was first predicted in Ref. \cite{Aceti:2015zva} and corroborated later experimentally by a BESIII experiment \cite{Ablikim:2015cob}. Similarly, in Ref. \cite{Aceti:2015pma} the $K \bar K \pi$ decay mode was studied and also found consistent with experiment \cite{Barberis:1997vf,Barberis:1998by}.
In the present work, we shall see that, as a consequence of the $K^* \bar K$ nature of the $f_1(1285)$, if we excite that state and go to higher energies where the $K^* \bar K$ can be produced, the tail of the $f_1(1285)$ propagator, together with the phase space for $K^* \bar K$ production, produce a peak around 1420 MeV with a width of about 60 MeV, that explains the experimental features observed for the $f_1(1420)$ resonance.

On the other hand, the triangle diagram with $K^* \bar K K$ intermediate states, with $\pi$ and $f_0(980)$ or $a_0(980)$ external products, develops a singularity at 1420 MeV, which is seen as a peak for $\pi f_0(980)$ or $\pi a_0(980)$ production. This was already suggested in Ref. \cite{Liu:2015taa} and shown explicitly in Refs. \cite{Ketzer:2015tqa,Aceti:2016yeb}, and provided a natural explanation of the COMPASS observation \cite{Adolph:2015pws} of a peak in the $\pi f_0(980)$ mode around 1420 MeV, which was interpreted as a new resonance, the ``$a_1(1420)$'' in Ref. \cite{Adolph:2015pws}. In Refs. \cite{Ketzer:2015tqa,Aceti:2016yeb} the peak was interpreted as a consequence of a triangle singularity associated to the decay mode of the $a_1(1260)$ into $K^* \bar K$, followed by $K^* \to K \pi$ and fusion of the $K \bar K$ to give the $f_0(980)$. The mechanism to produce the $\pi a_0(980)$ from the decay of the $f_1(1285)$ is identical to the one used in Ref. \cite{Aceti:2016yeb} to produce the $\pi f_0(980)$ from the decay of the $a_1(1260)$. Only the isospin combinations are different, but the singularity is tied to the masses of the particles and is independent of the internal degrees of freedom like the isospin. We shall see that, also in this case, a peak is produced around 1420 MeV in the $\pi a_0(980)$ decay mode of the $f_1(1285)$ which explains the features of the experiment of Ref. \cite{Barberis:1998by}.
The conclusion of all these observations is that the $f_1(1420)$ is not a genuine resonance, but the manifestation of the $K^* \bar K$ and $\pi a_0(980)$ decay modes of the $f_1(1285)$ resonance. This would go in line with the conclusions of Refs.  \cite{Ketzer:2015tqa,Aceti:2016yeb} that the ``$a_1(1420)$'' is not a genuine resonance, but the manifestation of the $\pi f_0(980)$ decay mode of the $a_1(1260)$ resonance.

\section{Formalism}

\begin{figure}[h!]
  \centering
  \includegraphics[width=0.4\textwidth]{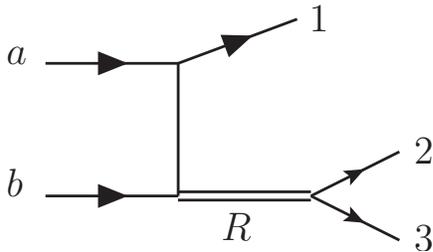}\\
  \caption{Diagrammatic representation of the process producing the resonance, observed in the decay channel $2+3$.}\label{Fig:1}
\end{figure}

The resonance $f_1(1420)$ is observed in very high energy collisions, which we depict in Fig. \ref{Fig:1}, and the resonance is observed in the invariant mass of particles 2 and 3.

In the Mandl and Shaw normalization of fermion fields \cite{MandlShaw:2010} we have for this process with three particles in the final state
\begin{equation}\label{Eq:1}
  \frac{\mathrm{d}^2\sigma}{\mathrm{\rm d} t \,\mathrm{d} M_{23} } = \frac{1}{32\, p_a^2 \, s}\, \prod(2m_F) \frac{1}{(2\pi)^3}\, \widetilde{p}_2 \overline{\sum}\sum|T|^2,
\end{equation}
where $m_F$ refers to the masses of the fermions, and the sum and average of $|T|^2$ is done over the polarizations of all particles involved. In Eq. (\ref{Eq:1}) $\widetilde{p}_2$ is the momentum of particle 2 in the rest frame of the $2+3$ system,
\begin{equation}\label{Eq:2}
\widetilde{p}_2 = \frac{\lambda^{1/2}(M_{23}^2,m_2^2,m_3^2)}{2M_{23}},
\end{equation}
and $M_{23}$ is the invariant mass of the $2+3$ system.
Fixing the Mandelstam variables $s$ and $t$, $s=(p_a + p_b)^2$, $t=(p_a - p_1)^2$, the $T$ matrix in Eq. (\ref{Eq:1}) will be of the type ($M_{\rm inv} \equiv M_{23}$):
\begin{equation}\label{Eq:3}
  T \equiv C \, \frac{1}{M_{\rm inv}^2 - M_R^2 + i M_R \Gamma_R} \, g_{R,23}\, ,
\end{equation}
where we have put a coupling $g_{R,23}$ for the resonance $R$ to the $2+3$ system. The width for the resonance going to $2+3$ is given in this case by
\begin{equation}\label{Eq:4}
  \Gamma_{R,23} = \frac{1}{8\pi} \frac{1}{M_{\rm inv}^2} \, \widetilde{p}_2 \,g_{R,23}^2 \, ,
\end{equation}
which allows one to write Eq. (\ref{Eq:1}) as
\begin{equation}\label{Eq:5}
  \frac{\mathrm{d}^2\sigma}{\mathrm{\rm d} t \,\mathrm{d} M_{\rm inv} } = \frac{\prod(2m_F)}{32\, p_a^2 \, s}\, \frac{1}{(2\pi)^3}\, C^2  \frac{8 \pi M_{\rm inv}^2 \Gamma_{R,23}}{|M_{\rm inv}^2 - M_R^2 + i M_R \Gamma_R|^2}.
\end{equation}

This equation is also good when we sum and average over polarizations of the particles in a more general case of $|T|^2$. The sum over polarizations of 2 and 3 will go into $\Gamma_{R,23}$ and the sum and average over polarizations of the other particles can be absorbed in the constant $C$ (for fixed $s$ and $t$). Eq. (\ref{Eq:5}) can also be used for any decay channel of the resonance, substituting $\Gamma_{R,23}$ by $\Gamma_i$ of the particular decay channel. Since $\sum \Gamma_i = \Gamma_R$, the total width of the resonance, the sum of Eq. (\ref{Eq:5}) over all decay channels, can be cast in terms of ${\rm Im}[1/(M_{\rm inv}^2 - M_R^2 + i M_R \Gamma_R)]$, which is a variant of the optical theorem.

\section{The $K^* \bar K$ channel}

Let us look at the process depicted in Fig. \ref{Fig:1} with 2, 3 being $K^* \bar K$. We shall investigate what happens when the resonance $R$ is the $f_1(1285)$. We can use directly Eq. (\ref{Eq:5}) replacing the coupling $g_{R,23}$ in Eq. (\ref{Eq:4}) by the $f_1, K^* \bar K$ coupling
\begin{equation}\label{Eq:6}
g_{R,23} \to g_{f_1, K^* \bar K} \; \vec{\epsilon}_{f_1}  \cdot  \vec{\epsilon}_{K^*} \, ,
\end{equation}
and summing and averaging over polarizations,
and we use $g_{f_1, K^* \bar K} = 7555$ MeV, from Ref. \cite{Roca:2005nm}, ignoring for the moment the isospin and $C$-parity structure. The $\Gamma_{f_1, K^* \bar K}$ width is then given by
\begin{equation}\label{Eq:7}
  \Gamma_{f_1, K^* \bar K} = \frac{1}{8\pi}\frac{1}{M_{\rm inv}^2} g_{f_1, K^* \bar K}^2 \, \widetilde{p}_{\bar K} \, \theta(M_{\rm inv} - m_{K^*} -m_K).
\end{equation}

\begin{figure}[h!]
  \centering
  \includegraphics[width=0.5\textwidth]{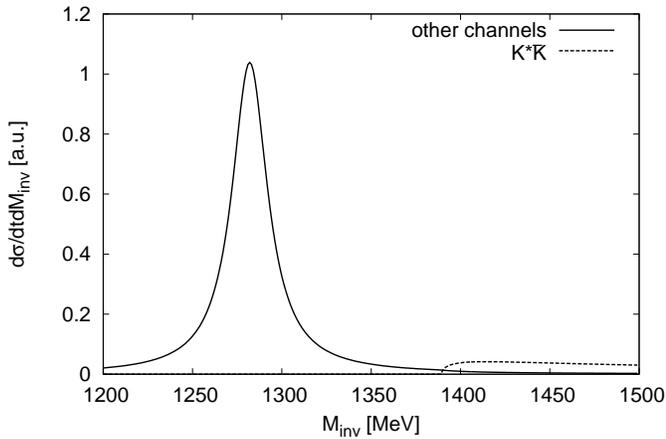}\\
  \caption{Production of $K^* \bar K$ induced by the excitation of the $f_1(1285)$ resonance. Dashed line: production of $f_1(1285)$ through the decay of $f_1(1285)$ into $K^* \bar K$. Solid line: through the decay of the $f_1(1285)$ into other channels.}\label{Fig:2}
\end{figure}

The $K^* \bar K$ production will begin at a threshold of 1383 MeV, 100 MeV above the nominal mass of the $f_1(1285)$. The results of Eq. (\ref{Eq:5}) are depicted in Fig. \ref{Fig:2}, where in addition we plot also the result using in the numerator the nominal width of the $f_1(1285)$, $\Gamma_{f_1} = 24.1$ MeV, which will account for the decays of the $f_1(1285)$ into other channels. In the denominator of Eq. (\ref{Eq:5}) we use
\begin{equation}\label{Eq:8}
  \Gamma_R = \Gamma_{f_1} + \Gamma_{f_1, K^* \bar K}.
\end{equation}

We observe in Fig. \ref{Fig:2} the typical threshold production of a channel. However, since the production is driven by the excitation of the $f_1(1285)$, one has two factors competing, a decreasing strength of the resonance as the energy increases, and an increasing phase space for the  $K^* \bar K$ production, and the product of these two factors confers the cross section a particular shape. Yet, we want to be more accurate here by taking into account that the $K^*$ will decay into $K \pi$ and the experimentalist will observe $K \pi \bar K$ at the end. This will also allow us to go below the threshold of $K^* \bar K$ production.

\begin{figure}[h!]
  \centering
  \includegraphics[width=0.4\textwidth]{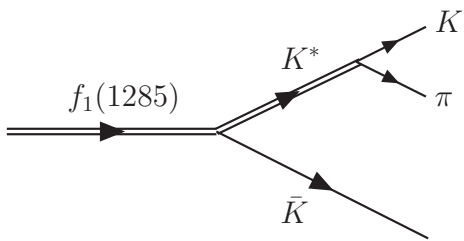}\\
  \caption{Decay diagram for $f_1(1285) \to K^* \bar K$ considering the $K \pi$ decay channel of the $K^*$.}\label{Fig:3}
\end{figure}

To account for the decay, we look into the diagram of Fig. \ref{Fig:3}.

The decay width of the $f_1(1285)$ for Fig. \ref{Fig:3} is given by
\begin{equation}\label{Eq:9}
   \Gamma_{f_1, K^* \bar K \to K \pi \bar K } \equiv
   \int \frac{\mathrm{d} \Gamma_{f_1, K \pi \bar K}}{\mathrm{d}m_{\rm inv}}\mathrm{d}m_{\rm inv},
\end{equation}
where $m_{\rm inv}$ stands for the $K \pi$ invariant mass, and
\begin{equation}\label{Eq:10}
   \frac{\mathrm{d} \Gamma_{f_1, K \pi \bar K}}{\mathrm{d}m_{\rm inv}} =
   \frac{1}{(2\pi)^3}\frac{1}{4M_{\rm inv}^2} \, p_{\bar K} \, \widetilde{p}_\pi \, \overline{\sum} \, \sum \, |t'|^2,
\end{equation}
where $p_{\bar K}$ is the $\bar K$ momentum in the $f_1(1285)$ rest frame and $\widetilde{p}_\pi$ the $\pi$ momentum in the $K \pi$ rest frame, and $t'$ contains now the $K^*$ propagator, the $f_1(1285) \to K^* K$ and $K^* \to K \pi$ couplings. There is no need to evaluate $t'$ explicitly since we can use the same step that led us from Eq. (\ref{Eq:1}) to Eq. (\ref{Eq:5}) and we can write
\begin{align}\label{Eq:11}
   \nonumber \frac{\mathrm{d} \Gamma_{f_1, K \pi \bar K}}{\mathrm{d}m_{\rm inv}} &=
   \frac{1}{(2\pi)^3}\frac{1}{4M_{\rm inv}^2} \, g_{f_1,K^* \bar K}^2 \, p_{\bar K}~~
   \\ &\times ~~
 \frac{8 \pi m_{\rm inv}^2 \Gamma_{K^*}(m_{\rm inv})}{|m_{\rm inv}^2 - m_{K^*}^2 + i m_{K^*} \Gamma_{K^*}(m_{\rm inv})|^2}\, ,
\end{align}
where
\begin{equation}\label{Eq:12}
  \Gamma_{K^*}(m_{\rm inv}) = \frac{m_{K^*}^2}{m_{\rm inv}^2} \, \frac{{\widetilde{p}_\pi}^{\,3}}{\widetilde{p}_\pi^{\,3}|_{\rm on}}\,\Gamma_{K^*}|_{\rm on}\, ,
\end{equation}
where $\Gamma_{K^*}|_{\rm on}=49.1$ MeV is the nominal width for the $K^*$, and
$\widetilde{p}_\pi$, $\widetilde{p}_\pi|_{\rm on}$ the pion momenta in the $K^*$ rest frame with
$K^*$ mass $m_{\rm inv}$ or $m_{K^*}$ respectively,
\begin{equation}\label{Eq:13}
\widetilde{p}_\pi = \frac{\lambda^{1/2}(m_{\rm inv}^2,m_K^2,m_\pi^2)}{2m_{\rm inv}}.
\end{equation}
In Eq. (\ref{Eq:12}) we have taken into account that the $K^* \to K \pi$ proceeds in $P$-wave.

\begin{figure}[h!]
  \centering
  \includegraphics[width=0.5\textwidth]{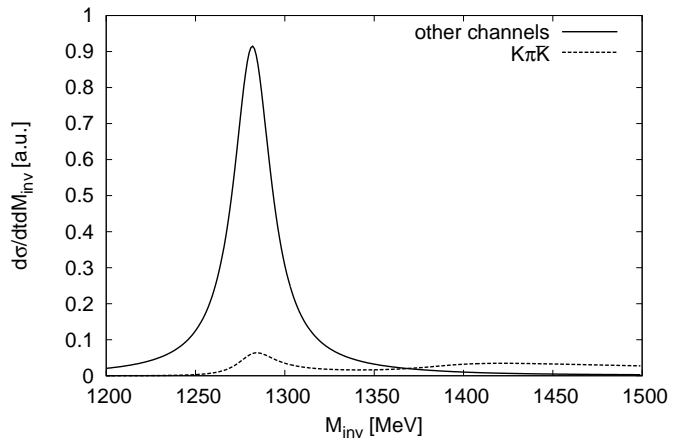}\\
  \caption{Production of $K \pi \bar K$ coming from $K^* \bar K$ decay of the $f_1(1285)$ considering the decay of $K^*$ into $K \pi$. The dashed line accounts for $K \pi \bar K$ production and the solid line accounts for the decay of the $f_1(1285)$ into other channels.}\label{Fig:4}
\end{figure}

In Fig. \ref{Fig:4} we plot the results for Eq. (\ref{Eq:5}) with $\Gamma_{R,23}$ replaced by Eqs. (\ref{Eq:9}) and (\ref{Eq:11}), and also when $\Gamma_{R,23}$ is replaced by $\Gamma_{f_1}$.
Fig. \ref{Fig:4} is very intuitive, we see a double peak structure. The first peak accounts for the standard $f_1(1285)$ decay into $K \pi \bar K$ observed with the shape of the $f_1(1285)$. The ratio of strengths at the peak of the dashed and solid lines provides the branching ratio of the $f_1(1285)$ into $K \pi \bar K$ channel. As we see, it is of the order of 8\%, the same value obtained in Ref. \cite{Aceti:2015pma} with a more elaborate model that we shall discuss below, and in agreement with experiment \cite{PDG}.

\begin{figure}[h!]
  \centering
  \includegraphics[width=0.5\textwidth]{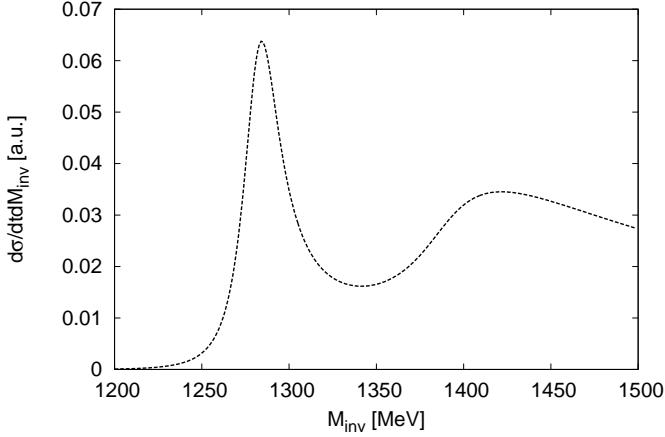}\\
  \caption{The $K\pi\bar K$ production cross section from Eq. \eqref{Eq:5} showing detail of the dashed line in Fig \ref{Fig:4}.}\label{Fig:4bis}
\end{figure}

Yet, what concerns us here is that the same mechanism produces a second peak around 1420 MeV as a consequence of the influence of the tail of the $f_1(1285)$ resonance and the increasing phase space for $K^* \bar K$ production. In Fig. \ref{Fig:4bis} we can see in more detail the two peaks corresponding to $K\pi\bar K$ production. By assuming a smooth background below the second peak, as an experimental analysis would do, we induce that there is a resonant-like structure peaking around 1420 MeV with a width of about 60 MeV, the features observed in experiment when talking about the $f_1(1420)$ resonance. Yet, we did not have to invoke a new resonance for this structure, which appears naturally and unavoidably from the decay of $f_1(1285)$ into $K^* \bar K \to K \pi \bar K$. This also explains why the $f_1(1420)$ resonance is seen in the $K^* \bar K$ (or $K \pi \bar K$) channel alone. In the next section we address the production of the $\pi a_0(980)$ channel.

\section{The $\pi a_0(980)$ decay mode of the $f_1(1285)$}

\begin{figure}[h!]
  \centering
  \includegraphics[width=0.4\textwidth]{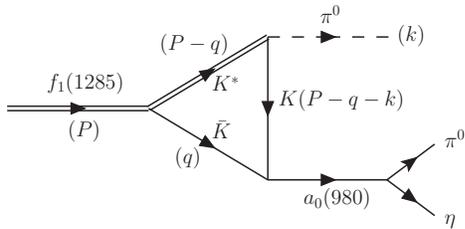}\\
  \caption{Triangle diagram leading to the production of $\pi a_0(980)$, the latter is observed in $\pi^0 \eta$. In brackets the momenta of the particles.}\label{Fig:5}
\end{figure}

This problem was also addressed in Ref. \cite{Aceti:2015zva} but at the peak of the $f_1(1285)$. Now we extend it to higher energies according to Eq. (\ref{Eq:5}), but replacing the width $\Gamma_{R,23}$ by $\Gamma_{f_1(1285),\pi a_0(980)}$. Following Ref. \cite{Aceti:2015zva} and sticking to the simplified $K^* \bar K$ decay of the former section, we have to look at the diagram of Fig. \ref{Fig:5}, where we consider the $a_0(980)$ decay into the $\pi^0 \eta$ channel.

The amplitude for the diagram of Fig. \ref{Fig:5} for the $f_1(1285)$ at rest ($\vec{P}=0$) is given by
\begin{equation}\label{Eq:14}
  t_{f_1, \pi \pi^0 \eta} = g_{f_1, K^* \bar K} \, g_{K^*, K \pi} \, t_T \, t_{K \bar K \to \pi^0 \eta} (\widetilde{m}_{\rm inv}),
\end{equation}
with $\widetilde{m}_{\rm inv}$ the $\pi^0 \eta$ invariant mass, and where $t_T$ stands for the triangle loop integral with three propagators,
\begin{widetext}
\begin{equation}\label{Eq:15}
  t_T = i \int \frac{\mathrm{d}^4q}{(2\pi)^4}\, \vec{\epsilon}_{f_1}  \cdot  \vec{\epsilon}_{K^*}\;
  \vec{\epsilon}_{K^*} \cdot (2\vec{k} + \vec{q}) \,
  \frac{1}{q^2 - m_K^2 +i\epsilon}\,
  \frac{1}{(P-q)^2 - m_{K^*}^2 +i m_{K^*} \Gamma_{K^*}}\,
  \frac{1}{(P-q-k)^2 - m_K^2 +i\epsilon} \, ,
\end{equation}
\end{widetext}
and $t_{K \bar K \to \pi^0 \eta}$ is evaluated using the chiral unitary approach of Ref. \cite{Oller:1997ti}, with the input used in the study of weak decays of $B$ and $D$ mesons in Refs. \cite{Liang:2014tia,Xie:2014tma}.

The integral of $q^0$ in Eq. (\ref{Eq:15}) can be done analytically and it is done in Ref. \cite{Aceti:2015zva} with the result

\begin{equation}\label{Eq:16}
  t_T = \widetilde{t}_T \, \vec{\epsilon}_{f_1} \cdot \vec{k} ,
\end{equation}
and

\begin{widetext}
\begin{align}\label{Eq:17}
 \nonumber \widetilde{t}_T &= \int \frac{\mathrm{d}^3q}{(2\pi)^3}
 \left(2 + \frac{\vec{k} \cdot \vec{q}}{\vec{k}^2} \right)\,
 \frac{1}{8\, \omega(q) \omega^{\prime}(q) \omega^*(q)}\,
  \frac{1}{k^0 -\omega^{\prime}(q) -\omega^{*}(q) +i\epsilon}\,
  \frac{1}{P^0 -\omega^*(q) -\omega(q) +i\epsilon}\\
&\times
\frac{2P^0 \omega(q) +2k^0\omega^{\prime}(q)
-2(\omega(q) +\omega^{\prime}(q))
(\omega(q) +\omega^{\prime}(q) +\omega^{*}(q))}{(P^0 -\omega(q) -\omega^{\prime}(q) -k^0+i\epsilon) (P^0+\omega(q)+ \omega^{\prime}(q) -k^0 -i\epsilon)}\, ,
\end{align}
\end{widetext}
with $\omega(q) = \sqrt{\vec{q}^{\,2} +m_K^2}$, $\omega^{\prime}(q) = \sqrt{(\vec{q}+\vec{k})^2 +m_K^2}$, $\omega^{*}(q) = \sqrt{\vec{q}^{\,2} +m_{K^*}^2}$ ,
(or $\sqrt{\vec{q}^{\,2} +m_{K^*}^2} - i \Gamma_{K^*}/2$ if the $K^*$ width is considered as in Eq. (\ref{Eq:15})).

The width $\Gamma_{f_1(1285),\pi\pi^0\eta}$ is given by
\begin{equation}\label{Eq:18}
\Gamma_{f_1,\pi\pi^0\eta}=\int{\frac{\mathrm{d}\Gamma_{f_1,\pi\pi^0\eta}}{\mathrm{d}\tilde{m}_{\rm inv}}\mathrm{d}\tilde{m}_{\rm inv}}\, ,
\end{equation}
with $\mathrm{d}\Gamma_{f_1,\pi\pi^0\eta}/\mathrm{d}\tilde{m}_{\rm inv}$ obtained using Eq. (\ref{Eq:10}) with $m_{\rm inv}\to \tilde{m}_{\rm inv}$ (of $\pi^0\eta$),
$p_{\bar K} \to p_{\pi}$,
$\tilde{p}_{\pi} \to \tilde{p}_{\eta}$ and
\begin{equation}\label{Eq:19}
\overline{\sum}\sum|t_{f_1,\pi\pi^0\eta}|^2=\frac{1}{3}\vec{k}^{\,2}|\tilde{t}_T|^2\,g_{f_1,K^*\bar K}^2\,g_{K^*,K\pi}^2\,|t_{K\bar K,\pi^0\eta}|^2\, .
\end{equation}

The value of $g_{K^*,K\pi}$ is taken such that
\begin{equation}
\Gamma_{K^*}=\frac{4}{3}\frac{1}{8\pi}\frac{1}{m_{K^*}^2}\tilde{p}_{\pi}^2g_{K^*,K\pi}^2
\end{equation}
gives the width of the $K^*$, $\Gamma_{K^*}=49.1$ MeV ($g_{K^*,K\pi}=5.5$ MeV).

\begin{figure}[h!]
  \centering
  \includegraphics[width=0.5\textwidth]{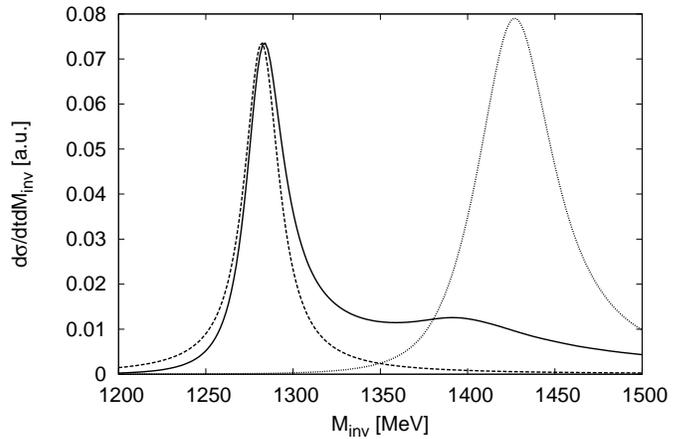}\\
  \caption{
  Differential cross section for $\pi a_0(980)\to\pi\pi^0\eta$ production induced by $f_1(1285)$ excitation. Solid line: considering the triangle diagram of Fig. \ref{Fig:5}. Dashed line: using Eq. (\ref{Eq:5}) with $\Gamma_R = \Gamma_{f_1(1285)} = 24.1$ MeV, $\Gamma_{R,23}$ a constant and normalizing the curves to the peak of the $f_1(1285)$ (this reflects the shape of the modulus square of the $f_1(1285)$ propagator). The dotted curve is what we would expect for $\pi a_0(980)\to\pi\pi^0\eta$ production from $f_1(1420)$ excitation assuming that the production rate of the $f_1(1285)$ and $f_1(1420)$ are the same. }\label{Fig:6}
\end{figure}

We should note a small modification in the integral with respect to Ref. \cite{Aceti:2015zva}. In this latter work a cutoff $\theta(q_{\rm max}-q)$ originated in the chiral unitary approach in the study of the $K\bar K\to \pi \eta$ amplitude was implemented, with $q_{\rm max}=600$ MeV \cite{Liang:2014tia,Xie:2014tma}. Since in those works the cutoff is implemented in the center of mass ($\rm CM$) frame, we make a boost of $q$ to the rest frame of the $\pi\eta$ system and implement the cutoff $\theta(q_{\rm max}-q_{\rm CM})$. We show the results obtained in Fig. \ref{Fig:6}.

As we can see in Fig. \ref{Fig:6}, there is a large strength in the cross section built around 1400 MeV induced by the $f_1(1285)$ excitation, which makes it very distinct from the usual shape of the Breit-Wigner distribution for the $f_1(1285)$. It is interesting to see that this cross section is remarkably similar to the one found in Ref. \cite{Barberis:1998by} (we shall come back to it in the next section to do a comparison with the data). In Ref. \cite{Barberis:1998by} no explanation was found for this extra strength and it was suggested that it should be the $\pi a_0(980)$ decay mode of the $f_1(1420)$. What we see here is that the peculiar shape of the cross section for this particular channel is a consequence of the triangle diagram of Fig. \ref{Fig:5}, the mechanism for   $\pi a_0(980)$ production from a resonance (the $f_1(1285)$) that is a bound state of $K^*\bar K$, or that couples strongly to $K^*\bar K$ for the purpose (the $f_1(1285)$). The unexpected large strength around 1400 MeV comes  from a singularity in the triangle diagram as we discuss in the next section. From that we can conclude that the strength found in this channel around 1400 MeV is not tied to the $f_1(1420)$ resonance but to the $f_1(1285)$.

There is one more argument that we can bring in favour of the former interpretation. Indeed, let us assume that the $f_1(1420)$ is a genuine resonance. If it decays into $K^*\bar K$ and this channel exhausts the width as found experimentally, we can get the coupling $g_{f_1(1420),K^*\bar K}$ by means of Eq. (\ref{Eq:4}) and we find
\begin{equation}\label{Eq:20}
g_{f_1(1420),K^*\bar K}=4256\ \rm MeV\, .
\end{equation}
With this coupling we can reevaluate the triangle diagram of Fig. \ref{Fig:5}, simply replacing the coupling $g_{f_1(1285),K^*\bar K}$ with the new one. Then we would use Eq. (\ref{Eq:5}) to get the cross section for $f_1(1420) \to \pi a_0(980)$, replacing the $f_1(1285)$ propagator by the one of the $f_1(1420)$. The other change might be the constant $C$, but this constant, appearing in the $a+b\to 1+R$ vertex (see Fig. \ref{Fig:1}) is related to the resonance production (irrespective of the decay channels).
Since one is dealing with high energies of the order of hundreds of GeV in these reactions, and the masses of the $f_1(1285)$ and $f_1(1420)$ are similar, on statistical grounds we should expect similar production rates for both resonances, and hence $C$ should be similar in the two cases.
Assuming $C$ is the same for both resonances, what we see in Fig. \ref{Fig:6} is a shape for $f_1(1420) \to \pi a_0(980)$ production very different from the one coming from the $f_1(1285)$ decay, which reflects a $f_1(1420)$ Breit-Wigner structure, and most importantly, with a very large strength, which is not seen in any experiment. Even this signal reduced by a factor five should be clearly seen experimentally, and on statistical grounds it is not easy to justify that the production of the $f_1(1420)$ should be reduced by a factor five with respect to that of the $f_1(1285)$.

There is yet another factor to note. When performing the previous calculation that involves the $f_1(1420)$ propagator we had to calculate the width of the $f_1(1420)$ into $\pi a_0(980)$. We find that (see next section for a more detailed evaluation)
\begin{equation}\label{Eq:21}
\frac{BR(f_1(1420)\to\pi a_0(980))}{BR(f_1(1420)\to K^* \bar K)}\simeq 0.17\, .
\end{equation}
This is a fraction that could not be missed and has not been found in any experiment where this decay mode has been searched for. Only in Ref. \cite{Barberis:1998by} a 5\% ratio was invoked by guessing that the shoulder seen in that decay mode around 1400 MeV was due to this resonance, but we have given a different interpretation for this feature.

It is interesting to recall here that the triangle singularity of Fig. \ref{Fig:5} also shows up in the decay of the $\eta(1405) \to \pi a_0(980)$ which was studied in Refs. \cite{Wu:2011yx,Aceti:2012dj}, together with the isospin violating $\eta(1405)\to \pi f_0(980)$ decay. This latter decay was abnormally enhanced due to the triangle singularity. In a follow up of Ref. \cite{Wu:2011yx} in Ref. \cite{Wu:2012pg} the idea is retaken and applied to study the decay of $J/\psi \to \gamma \eta(1405)(\eta(1475))$ with $\eta(1405) \to K \bar K \pi$, $\eta \pi \pi$ and $3\pi$. In this work the possible contribution of the $f_1(1420)$ in addition to the $\eta(1405)$ was discussed. The $f_1(1420)$ was assumed to be a regular resonance and the triangle singularity enhanced some decay modes, in spite of which its contribution relative to that of the $\eta(1405)$ was found small in the radiative $J/\psi$ decay. Although we find that the $f_1(1420)$ is not a genuine resonance, we have seen that there is indeed strength in the $K \pi \bar K$ and $\pi a_0(980)$ channels in the 1420 MeV region from the decay of the $f_1(1285)$. A reanalysis of the BESIII experimental data \cite{BESIII:2012aa} from this new perspective would be most interesting.

\section{The singularity of the triangle diagram}

The shoulder for the $f_1(1285)$ decay to $\pi\pi^0\eta$ around 1400 MeV has its origin in the singularity that the triangle diagram of Fig. \ref{Fig:5} develops at 1420 MeV. Triangular singularities were studied by Landau in Ref. \cite{Landau:1959fi} and stem from triangle diagrams when  the intermediate particles are all placed on-shell and the momenta are parallel ($-\vec{q}$ and $\vec{k}$ of Fig. \ref{Fig:5} go in the same direction). There is also a further constraint  which is that the mechanism can lead to a classical process with $R\to K^*\bar K$; $K^*\to K\pi$; $K\bar K\to a_0(980)$, which is stated by the Coleman-Norton theorem \cite{Coleman:1965xm}. A pedagogical and practically easy way to visualize it, and see where the triangular singularity appears, is shown in Ref. \cite{Bayar:2016ftu} (see Eq. (18) of that paper).

In the present case we can see the origin of the singularities by looking at the expression of Eq. (\ref{Eq:17}). The two factors in the denominator that develop singularities are
\begin{equation}\label{Eq:22}
D_1=P^0-\omega^{*}(\vec{q}\,)-\omega(\vec{q}\,)+i\epsilon
\end{equation}
and
\begin{equation}\label{Eq:23}
D_2=P^0-k^0-\omega(\vec{q}\,)-\omega(\vec{q}+\vec{k}\,)+i\epsilon\, .
\end{equation}

\begin{figure}[h!]
  \centering
  \includegraphics[width=0.5\textwidth]{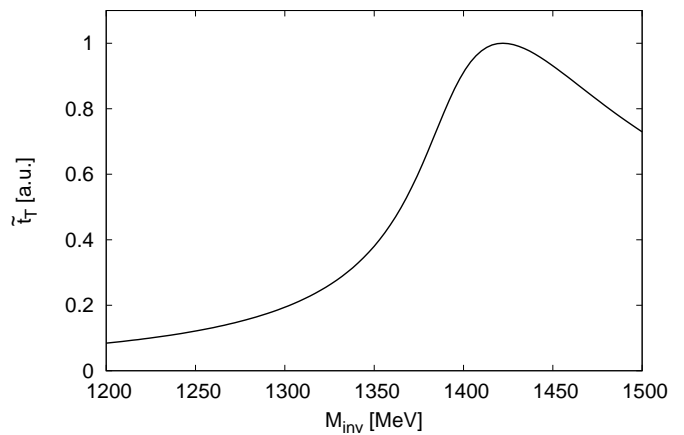}\\
  \caption{
  Results for the singular diagram, $|\tilde{t}_T|^2$, of Eq. (\ref{Eq:17}).}\label{Fig:7}
\end{figure}

When they are zero with $\vec q$ and $\vec k$ in opposite directions, one gets a pole when $D_1=0$ at $q_{\rm on}+i\epsilon$ and two poles when $D_2=0$, at $q_{a_+}+i\epsilon$ and $q_{a_-}-i\epsilon$. Then, when $q_{\rm on}=q_{a_-}$ two poles appear in opposite sides of the real axis at the same energy and the integral path cannot be deformed to avoid the singularity, which thus shows up in the result of the integral. If the width of the $K^*$ is considered, then $q_{\rm on}+i\epsilon\to q_{\rm on}+i\Gamma_{K^*}/2$ and the singularity turns into a peak. If we apply $q_{\rm on}=q_{a_-}$ (see explicit formulae in Ref. \cite{Bayar:2016ftu}) we find the singularity when the incoming energy in the triangle diagram is 1420 MeV. To show it, we plot in Fig. \ref{Fig:7} the result for $|\tilde{t}_T|^2$ of Eq. (\ref{Eq:17}). In Fig. \ref{Fig:7} we see how an original singularity becomes a broad peak when $\Gamma_{K^*}\neq 0$.
This softened singularity, together with the propagator of the $f_1(1285)$, is what gives rise to the broad shoulder of the $f_1(1285)\to \pi a_0(980)$ in Fig. \ref{Fig:6}.

\section{Detailed evaluation with the $I=0$ and $C=+$ parity of the $f_1(1285)$}

\begin{figure}[h!]
  \centering
  \includegraphics[width=0.5\textwidth]{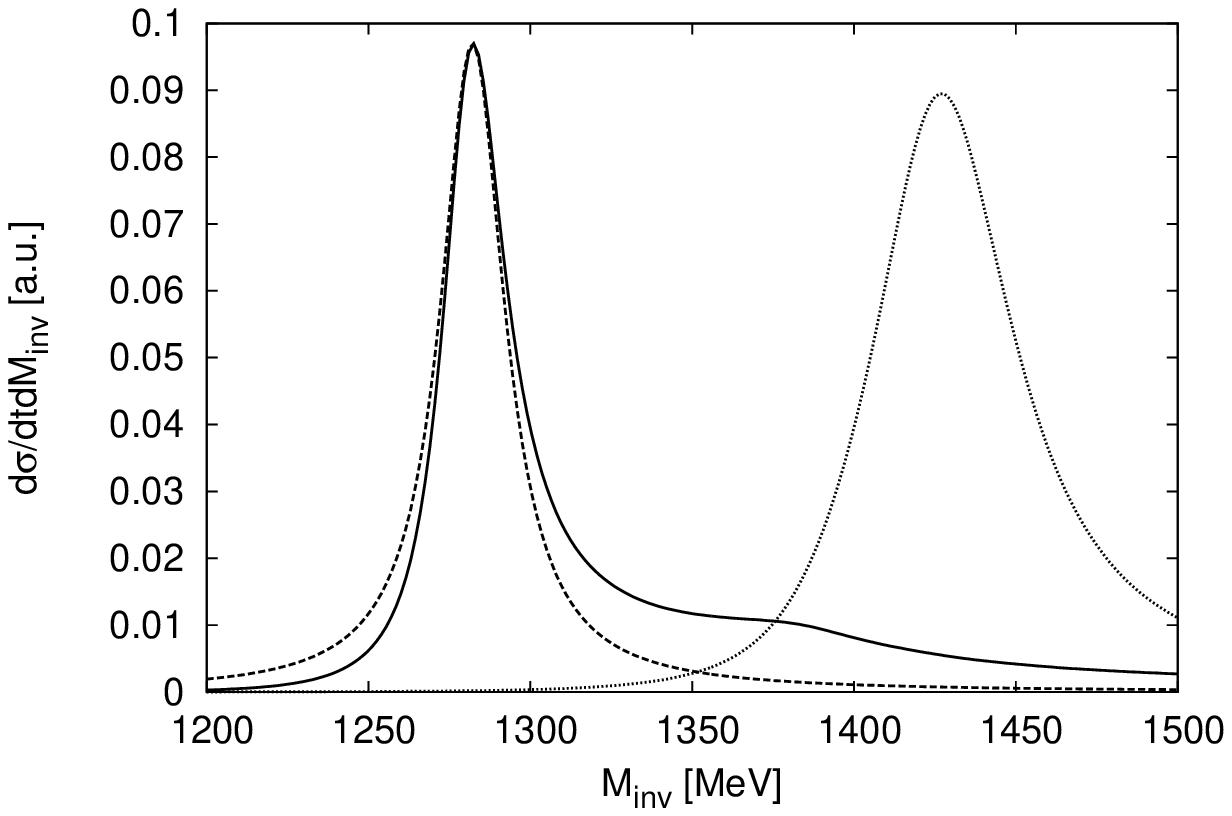}\\
  \caption{The differential cross section for the decay $f_1(1285)\to\pi a_0(980)$ with the full wave function of Eq. (\ref{Eq:24}).}\label{Fig:8}
\end{figure}

\begin{figure}[h!]
  \centering
  \includegraphics[width=0.5\textwidth]{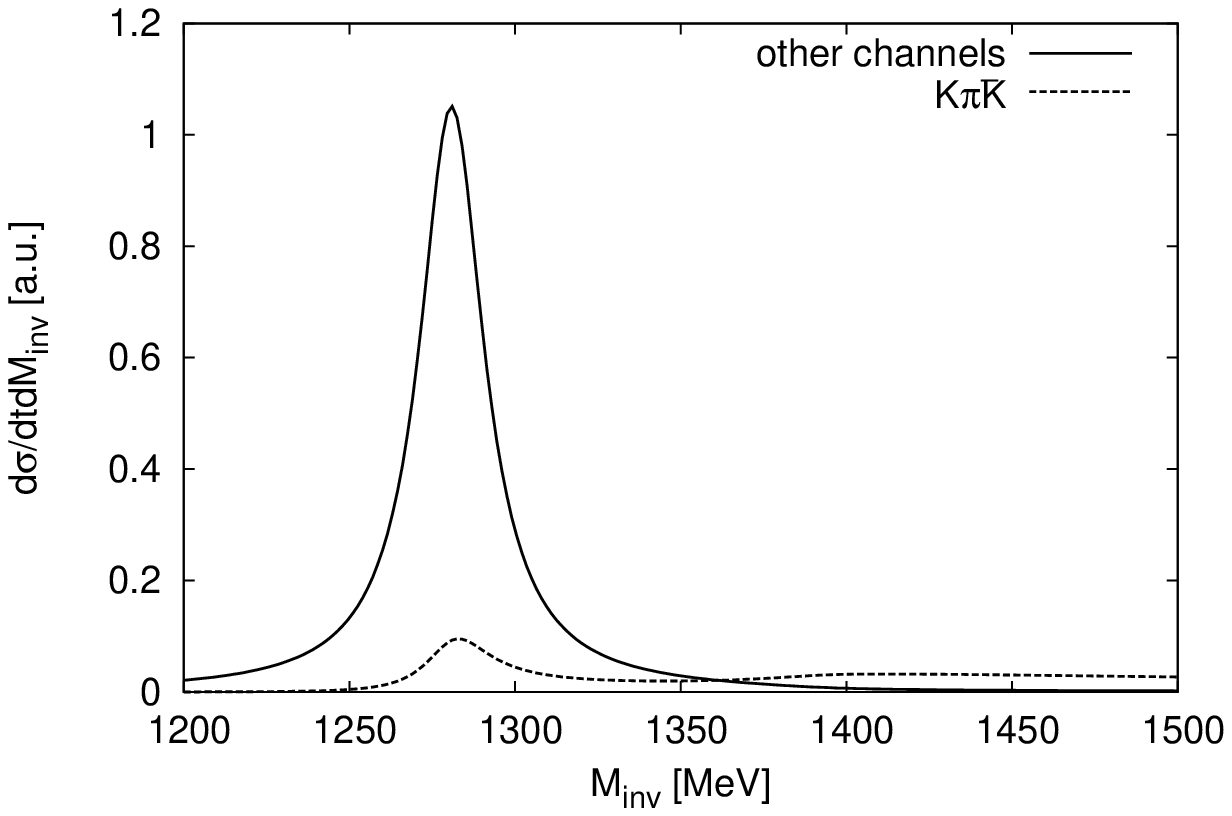}\\
  \caption{The differential cross section for the decay $f_1(1285)\to K\pi\bar K$ with the full wave function of Eq. (\ref{Eq:24}).}\label{Fig:9}
\end{figure}

\begin{figure}[h!]
  \centering
  \includegraphics[width=0.5\textwidth]{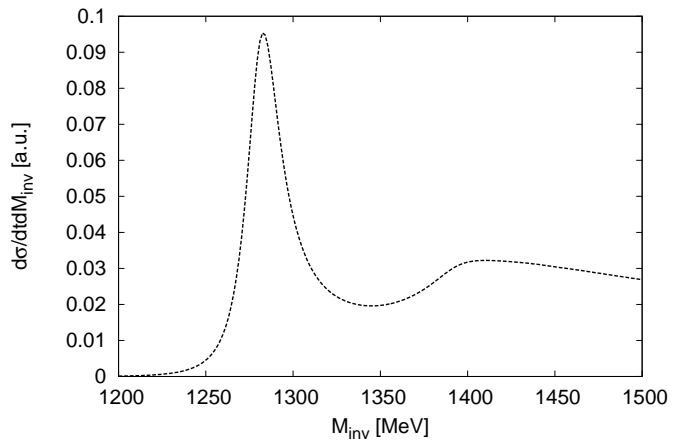}\\
  \caption{Same as Fig. \ref{Fig:4bis}  with the complete wave function of Eq. \eqref{Eq:24}.}\label{Fig:9bis}
\end{figure}

So far we did not pay attention to the isospin and $C$-parity structure of the $f_1(1285)$ and $f_1(1420)$, but the shapes and relative weights of the cross sections  are well evaluated with the previous formalism. The wave function for the $f_1(1285)$ is given by
\begin{equation}\label{Eq:24}
|f_1(1285)\rangle=-\frac{1}{2}(K^{*+}K^-+K^{*0}\bar{K}^{0}-K^{*-}K^+-\bar{K}^{*0}K^0)
\end{equation}
and then everything is identical to what have been done so far, except that one has four diagrams. The evaluation of the widths for $f_1(1285)\to \pi a_0(980)$ and $f_1(1285)\to \pi K \bar K$ at the peak of the $f_1(1285)$ are done in Refs. \cite{Aceti:2015zva} and \cite{Aceti:2015pma},  respectively. All that must be done is to perform the same evaluation as a function of $M_{\rm inv}$ (for the $f_1(1285)$ at higher energies), and implement the $f_1(1285)$ propagator in Eq. (\ref{Eq:5}). The results are shown in Figs. \ref{Fig:8}, \ref{Fig:9} and \ref{Fig:9bis}, which should be compared with Figs. \ref{Fig:6}, \ref{Fig:4} and \ref{Fig:4bis}, respectively. As we can see, not only the shapes but the absolute numbers are about the same as with the simplified wave function.
  Moreover, in Fig. \ref{Fig:9} a peak develops for the $f_1(1285)\to \pi K\bar K$ decay around 1400 MeV (see Fig. \ref{Fig:9bis} for more details). The peak and width of the distribution around this energy are in fair agreement with the mass of 1420 MeV and width of about 55 MeV quoted in the PDG \cite{PDG}.

We have taken advantage to make a more refined evaluation. Indeed, in Eq. \eqref{Eq:5} we have  $\Gamma_R\equiv\Gamma_{f_1(1285)}$ in the denominator. Since we are evaluating the partial decay width into $K\pi\bar K$ and $\pi a_0(980)$ we write
\begin{eqnarray}
\nonumber&\Gamma_R=\Gamma_{f_1(1285)}-\Gamma_{f_1(1285)\to K\pi\bar K}|_{\rm on}-\Gamma_{f_1(1285)\to \pi a_0(980)}|_{\rm on}
\\&+\Gamma_{f_1(1285)\to K\pi\bar K}(M_{\rm inv})+\Gamma_{f_1(1285)\to \pi a_0(980)}(M_{\rm inv})\, ,
\end{eqnarray}
which guarantees that on shell, $M_{\rm inv}=1285$ MeV, $\Gamma_{R}=24.1$ MeV. The values that we obtain are $\Gamma_{f_1(1285)\to K\pi\bar K}|_{\rm on}/\Gamma_{f1(1285)}=0.095$, $\Gamma_{f_1(1285)\to \pi a_0(980)}|_{\rm on}/\Gamma_{f1(1285)}=0.275$.

\begin{figure}[h!]
  \centering
  \includegraphics[width=0.5\textwidth]{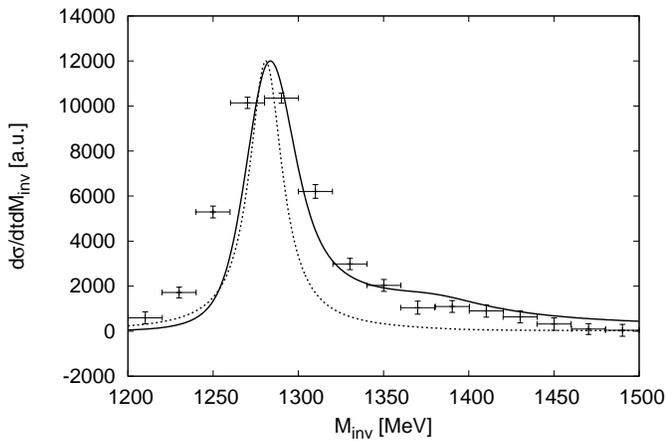}\\
  \caption{Comparison of the results of Fig. \ref{Fig:8}, convoluted with a resolution of 20 MeV, with the experimental results of \cite{Barberis:1998by}.}\label{Fig:10}
\end{figure}

We also make a more detailed comparison with the results of Ref. \cite{Barberis:1998by}. In Fig. \ref{Fig:10} we show our results folded with a resolution of 20 MeV to facilitate comparison with the experimental numbers. We normalize the results approximately to  the peak of the experimental distribution. We can see that the agreement with experiment is fair.

\section{Conclusions}

We have carried out a study of the production of the $f_1(1285)$ and decay into $\pi a_0(980)$ and $K^*\bar K$ modes. We have studied the cross sections as functions of the $f_1(1285)$ mass, $M_{\rm inv}$, up to 1500 MeV and we have observed two relevant features:
\begin{itemize}
\item[1)] The $K^*\bar K$ mode (allowing the $K^* \to K\pi$ decay) has two peaks as a function of $M_{\rm inv}$, one at the $f_1(1285)$ mass and the other one at about 1420 MeV, this latter one with a width of about 60 MeV.
\item[2)] The $\pi a_0(980)$ mode has a peak at 1285 MeV and a broad shoulder around 1400 MeV, which comes from a triangle singularity involving $K^* K \bar K$ as intermediate states, and tied to the nature of the $f_1(1285)$ as a $K^* \bar K$ molecule, a sufficient although not necessary condition, since what matters is that the $f_1(1285)$ couples to $K^* \bar K$ and this is known experimentally from the $K \pi \bar K $ decay mode. The combination of the tail of the $f_1(1285)$ with the increased phase space for the $K^* \bar K$ production is the reason for this second peak.
\end{itemize}

The two features described above are the experimental facts in which the $f_1(1420)$ was accepted as a resonance, but we have shown that they are consequence of the decay modes of the $f_1(1285)$ and one does not have to introduce any new resonance to account for these facts. The absence of the $\pi a_0(980)$ decay mode of the $f_1(1420)$ in all but one experiment \cite{Barberis:1998by}, is a fact that we have exploited here, because if it were a resonance which decays mostly into $K^*\bar K$, it would unavoidably have a width into $\pi a_0(980)$ of the order of 17\%, which has not been observed. The 5\% $\pi a_0(980)$ decay mode attributed to the $f_1(1420)$ in Ref. \cite{Barberis:1998by} was a guess based on the lack of any other interpretation of the shoulder found for this mode around 1400 MeV in the study of the decay of the $f_1(1285)$. We found a natural explanation for this broad peak which then does not require the existence of the $f_1(1420)$.

Altogether, our study leads us to the unavoidable conclusion that the $f_1(1420)$ is not a resonance but simply the manifestation of the $K^*\bar K$ and $\pi a_0(980)$ decay modes of the $f_1(1285)$ around 1420 MeV.

\section*{Acknowledgments}

One of us, V. R. D. wishes to acknowledge the support from the Programa Santiago Grisol\'ia of Generalitat Valenciana (Exp. GRISOLIA/2015/005).
One of us, E. O. wishes to acknowledge the support from the Chinese Academy of Science in the Program of Visiting Professorship for Senior International Scientists (Grant No. 2013T2J0012).
This work is partly supported by the National Natural Science Foundation of China under Grants No. 11565007, No. 11547307.
This work is also partly supported by the Spanish Ministerio de Economia y Competitividad and European FEDER funds under the contract number FIS2011-28853-C02-01, FIS2011-28853-C02-02, FIS2014-57026-REDT, FIS2014-51948-C2-1-P, and FIS2014-51948-C2-2-P, and the Generalitat Valenciana in the program Prometeo II-2014/068.

\clearpage

\bibliographystyle{plain}

\end{document}